\documentstyle[epsf,prl,twocolumn,aps]{revtex}
\begin{document}
\title{  Information about   the Integer Quantum Hall Transition Extracted 
 from the Autocorrelation Function of Spectral
 Determinants }
\author{Stefan Kettemann(1), Alexei Tsvelik(2)}
\address{(1) Max- Planck Institut f\" ur Physik Komplexer Systeme\\
 Noethnitzerstr. 38,
 01187 Dresden, Germany,\\
 (2) Department of Physics, University of Oxford, 1 Keble Road, Oxford, 
OX1 3NP, UK}
\date{\today }
\maketitle

\begin{abstract}
The  autocorrelation function of  spectral determinants (ASD) 
 is used to probe the sensitivity  
of a two-dimensional disordered  electron gas to the system's size $L$. 
  For weak magnetic fields the ASD is shown to depend only
trivially on $L$ which is a strong indication that all states are localized.
 From the nontrivial dependence of ASD on  $L$ at $L \rightarrow \infty$ 
 for strong magnetic fields 
at $ \sigma_{xy} = 1/2 $
we deduce the existence of critical wave functions. 
Pacs- numbers:  73.40.Hm,72.15.Rn,73.23.-b
\end{abstract}
  
 In the pioneering works on localization 
\onlinecite{anderson},
\onlinecite{mott} it was discovered  that an arbitrarily weak 
 disorder 
 leads to localization  in one dimension.  
   Both the scaling picture of Abrahams et al.\cite{ab}
 and the renormalization group
 treatment \cite{weg}, \cite{ef},
 \cite{elk}, \cite{larkin}
 of the Anderson model
 predicted a second order metal-insulator 
transition in dimensions larger than two.
 The dimension two has been  found to be
 marginal. For noninteracting electrons in a random 
 potential at no or weak magnetic field  no extended states
 and thus  no metal-insulator transition 
 has been found by  analytical and numerical methods\cite{hucke}. 
 
 However, Levine, Libby and Pruisken 
 noticed the presence of a
 topological term in 
the field theoretical formulation of the problem  (the nonlinear (NL) sigma 
 model)
 and argued that this may result in the existence of extended
 states in a strong magnetic field\cite{pruisken}.
 They suggested that these extended states  could explain 
 the  transition  
  between plateaus in the Hall conductance,
 where a finite longitudinal conductance is observed experimentally
\cite{klitzing}. 
  There is a numerical evidence for existence of extended 
 states in the  center
 of the Landau band\cite{hucke}. Since their wave functions 
  decay like a power law, these states have been called critical or
 prelocalized.

  In this paper we use the autocorrelation function
 of spectral determinants(ASD) to extract 
information about existence of critical states. 
The ASD is defined as follows: 
\begin{equation}
C(\omega) = \frac{\bar{C}(\omega)}{\bar{C}(0)},
\end{equation}
where
\begin{equation}
\bar{C}(\omega) = < det( E + \frac{1}{2}\omega - H) det ( E- \frac{1}{2}\omega -H ) >.
\end{equation}
   $H$ is the Hamiltonian of the  system
 under consideration, 
 and $E$ a central energy. 
      
  The  ASD  was used first    
 to characterize the spectrum of nonintegrable 
 quantum billiards\cite{and,haake,us}.
  Recently, 	a mesoscopic disordered metal wire 
 was studied with the ASD and  was shown to contain 
 information on the Anderson localization
 in such a wire\cite{me}.

 Despite  ASD being a non-self-averaging quantity, and  not  a generating 
 function for any physical observable,
 one can extract 
from it valuable qualitative information.  Namely, if there are 
extended states near the energy $E$, one expects the ASD to be sensitive 
to the system's size. Being more precise, in this case 
 we expect it to depend on $\omega^2 L^{\eta}$, where $\eta$ is a 
 constant, larger than the dimension $d$. 
To the contrary,  
if  all states are localized then the size dependence of the ASD
 can only be the trivial 
\begin{equation}     \label{trivial} 
 C = exp ( - \mbox{const} \tilde{V} \omega^2 ),
\end{equation}
 where $\tilde{V} = a L^d$ is the total volume of the sample,
 thus for a quasi- two- dimensional film $a$ is its thickness,
 and $d=2$.
 This can be proven using a
simple model used in Refs. 
{\onlinecite{sivan},\onlinecite{aronov}}:
 If all states are localized with an average localization length $L_c$, 
the  system  
can be divided into $(L/L_c)^d$ localization volumes.
 The levels in each localization volume obey Wigner- Dyson statistics
\cite{wigner}, \cite{dyson} and
  have an average level spacing of $ \Delta_{c} = 1/( \nu a L_c^d)$,
 where $\nu$ is the average density of states.
 The levels in different localization volumes  
 are almost uncorrelated. The levels of the total sample  have a  spacing
 $\Delta = 1/(\nu a L^d)$.
  Taking all the $ N_c = \Delta_c/\Delta$ levels of the wire 
in an energy interval $\Delta_{c}$, 
 they should be  uncorrelated. With 
 a Gaussian  distribution of width 
  $\Delta_c$ we  get 
 $ C( \omega ) = \exp [ - (\pi/2)  \omega^2/(\Delta \Delta_c)]$
 which  agrees indeed with Eq. (~\ref{trivial}~).

Here, we will use the ASD to address the question of 
 localization in 2D and the integer quantum Hall transition.
  It  allows us to answer the question 
 if there are extended states, and where in the 
 spectrum they are located.
 
As Hamiltonian we take  the Anderson model,
\begin{equation}
H= ({\bf p}- q/c {\bf A} )^2/(2m) + V({\bf x}),
\end{equation}
 where $q$ is the electron charge, c the velocity of light,
 and ${\bf A}$ the vector potential due to an external magnetic   field 
${\bf B}$.  
$V({\bf x})$ is a Gaussian distributed random function
\begin{equation}
< V({\bf x}) > = 0, < V({\bf x}) V({\bf x'}) > = \frac{\Delta}{ \tau} 
\frac{\delta ({\bf x} - {\bf x'})}{2 \pi a L^2},  
\end{equation}
 which models randomly distributed, uncorrelated impurities in the conductor.
 
When the time reversal symmetry is fully broken by an external 
 magnetic field,
 the ASD can  be represented by a functional integral over 
 two- component Grassmann fields. This representation is invariant under
 $U(2) $- transformations of these Grassmann fields.
  Averaging over impurities, the resulting interacting theory can be 
decoupled by introducing a Gaussian integral over $2\times 2$ 
 matrices. 
   Integrating over the Grassmann fields, 
  doing a saddle point approximation  and
 integrating over longitudinal (massive) modes 
 one obtains the ASD  
 for a quasi- two- dimensional disordered film of  
size $L$ and thickness $a$,
  for $\omega \ll 1/\tau$ as a functional integral over the transverse
 modes, see Refs. \onlinecite{pruisken2}, \onlinecite{weiden}: 
\begin{equation}
\bar{C} ( \omega ) = \int \prod_{\bf x} d Q ({\bf x})  \exp ( - F[Q] ),
\end{equation}
where 
\begin{eqnarray}\label{free}
F[Q] &=&  \frac{1 }{8} \sigma_{x x} (E)
 \int d^2 {\bf x} Tr  ( {\bf {\bf \nabla}} Q({\bf x}))^2
\nonumber \\
&+& \frac{\pi}{2} i \frac{\omega}{\Delta}
\int \frac{d^2 {\bf x}}{ L^2} Tr \sigma_3  Q({\bf x})
\nonumber \\
& - & \frac{1}{16} \sigma_{x y}(E) \int d^2 {\bf x} 
 Tr Q [\nabla_x Q, \nabla_y Q]
\end{eqnarray}
 Here, $\sigma_3$ is the diagonal Pauli matrix and 
 $\sigma_{xx}(E)$ and $\sigma_{xy}(E)$ are the {\it bare} longitudinal and 
 Hall conductance, in units of $e^2/h$,
 of the disordered conductor on energy $E$. 
 The transverse fluctuations are 
 restricted by
 $Q^2 = 1$, which reduces the manifold of Q to $U(2)/( U(1) \times U(1) ) 
 = SU(2)/U(1) = S_2$, the space of points on a sphere.
 This is at $\omega =0$  
 the free energy of the 
 two- dimensional  O(3)- NL sigma model
 with a topological term.

  The representation  of the matrix Q can be chosen as, 
\begin{equation}
Q = U Q_c \bar{U},
\end{equation}
 where
\begin{equation}
 Q_c = \left( \begin{array}{cc} \cos \theta & i \sin \theta
 \\  - i \sin \theta & - \cos \theta  \end{array} \right),
  U = \left( \begin{array}{cc}  \exp ( i \varphi/2 ) & 0 \\
 0 & \exp ( - i \varphi/2 ) \end{array} \right).
\end{equation}

 Thus,  we obtain:
\begin{equation}
 \bar{C} ( \omega ) = \int \prod_{\bf{x} } ( d \theta d \varphi \sin \theta )
\exp ( - F[ \theta, \varphi ] ),
\end{equation}
 where
\begin{eqnarray}\label{sigmod}
&& F[ {\bf n} ] = \frac{1}{4} \sigma_{xx} (E) \int d^2 x 
 \sum_{\mu =1}^2  ( \partial_{\mu} {\bf n} )^2
\nonumber \\ && 
 - \frac{i}{2} \sigma_{x y} (E) \int d^2 x {\bf n} ( \partial_x {\bf n}
 \times \partial_y {\bf n} )
  + i \pi \frac{ \omega}{\Delta} 
 \int d^2 x n_3.
\end{eqnarray}
 which is written in terms of the unit vector on a sphere,
 $ {\bf n} = ( \sin \theta \cos \varphi , \sin \theta
 \sin \varphi, \cos \theta ) $.

 At $\omega =0$  the model is integrable for
 $\sigma_{xy} = 0$ \cite{wiegmann} and $\sigma_{xy} = 1/2$
 \cite{zamzam}. The 
 exact solution  of the corresponding
(1+1)-dimensional model is known at these points. 
In the first case the spectrum  contains a spectral gap giving 
a finite correlation length in the Euclidean version of the model: 
$\xi_c \sim
\lambda \sigma_{xx}\exp( \pi \sigma_{xx})$\cite{wiegmann,amit}. 
This scale differs from
the localization length 
for unitary disorder  
$L_c \sim \lambda\exp( \mbox{const~ }\sigma_{xx}^2)$. Such discrepancy 
may be  connected with   ASD being a non-self-averaging 
quantity \cite{boris}.

 Let our system be
 defined on a square $L_x\times L_y$ with periodic boundary conditions.
 We can identify the $x$-direction 
with the Matsubara time and treat our sigma model as an Euclidean field theory at temperature $T = 1/L_x$. 
The ASD can be written as a partition function
\begin{equation}
C( \omega ) = Tr \exp ( - L_x  H ), 
\end{equation}
  where $H$ is the Hamiltonian on a circle of circumference $L_y$. 
 Thus, for $ L_x,L_y \rightarrow \infty$ one gets 
\begin{equation}
C(\omega)/C(0) = \exp(- L_xL_y{\cal F}(\omega/\omega_0;\xi_c/L_y))
\end{equation}
where ${\cal F}$ is the $\omega$-dependent part of the 
ground state energy density 
of the NL sigma model. In the limit $L_y >> \xi_c$ the 
$L$-dependence disappears from ${\cal F}$. 
In the  remaining form the system's size enters trivially: 
\begin{equation}
C(\omega)/C(0) = \exp(- L_xL_y{\cal F}(\omega/\omega_0))
\end{equation}
as one would expect for a bunch of localized levels. 
Moreover, at $\omega << \omega_0$ the $\omega n_3$ perturbation 
is non-singular and one can expand the free energy in $\omega$ to get ${\cal F}$$(\omega/\omega_0) \sim (\omega/\omega_0)^2$ (compare with (\ref{trivial})).

 For $\sigma_{xy} = 1/2$ the spectrum is gapless and at the critical
 point the model is equivalent to the level $k = 1$ SU(2)
 Wess-Zumino-Novikov-Witten  model (WZNW1) perturbed by the marginal 
current-current interaction \cite{zamzam}. This model is 
in the same universality class as the spin-1/2 isotropic 
Heisenberg antiferromagnetic chain. 
Since the
topological term does not change the classical equations of motion and 
therefore does not contribute to the perturbation theory
expansion in powers of $\sigma_{xx}^{-1}$,  the above result is
 non-perturbative. The  WZNW1- model  can be
 represented as a Gaussian model with a particular set of scaling dimensions:
\begin{equation}
S = \int d^2 x[\frac{1}{2}(\nabla\Phi)^2 + gJ^a\bar J^a]\label{Gauss}
\end{equation}
where $J^a, \bar J^a$ are currents satisfying the level 
$k = 1$ Kac-Moody algebra. The quantity $\xi_c$ serves as the ultraviolet
cut-off for theory (\ref{Gauss}) so that this model is
valid only for  distances greater than $\xi_c$. The coupling $g$ is 
 proportional to 
$\sigma_{xx} - \sigma^*$. 
At the critical point the operator 
$Tr\sigma_3 Q = 2 \cos \theta$  becomes 
 $F\cos\sqrt{2\pi}\Phi$, where $F$ is a prefactor whose exact value
is currently unknown. Hence  the equivalent
representation of the model (\ref{sigmod}) at distances greater than
$\xi_c$ and $\omega/\omega_0 << 1$ is
\begin{equation}\label{jump}
S = \frac{1}{2}\int d^2x[(\nabla\Phi)^2 + \pi i 
\frac{\omega}{\omega_0} F \cos\sqrt{2\pi}\Phi  + gJ^a\bar J^a] \label{SG}
\end{equation}  
At $g > 0$ the last term is marginally irrelevant and can be neglected. 
After that we get  the sine-Gordon model which  is exactly solvable though
most of the  results have been obtained 
for real i$\omega$. 
The reality of
$\omega$  gives rise to interesting new features (see \cite{fsz}). 
The quantity $\omega_0$ is proportional to the inverse correlation
length of the original sigma model squared $\omega_0 \sim D/\xi_c^2$. 
Since the new theory is defined on distances greater than $\xi_c$,  
 the  correlation
 functions of the exponentials are normalized as
\[
<\exp(i\beta\Phi(x))\exp(-i\beta\Phi(x))> 
\sim \left(\frac{\xi_c}{|x|}\right)^{\beta^2/2/\pi}
\]
For the $n_3$-operator at the critical
point $\beta = \sqrt{2 \pi}$.
 From  the poor man's scaling
one gets then:
\begin{equation}
C(\omega)/C(0) = \exp[- f(\omega^2L^3)]\label{C}
\end{equation}
where $f(x) \sim x$ at $x << x_0$,
 $x_0 = \omega_0^2 \xi_c^3$ (here we put $L_x = L_y = L$).
 This is different  from the trivial size dependence Eq. ( \ref{trivial} )
 one obtains when all states are localized.  
Thus this ASD remains sensitive to the system's size at 
any $\omega << \omega_0$, which 
 implies 
 the existence of extended states.

 The 
renormalized coupling constant
  of the O(3) sigma model discussed in this work
 is not directly  related to the physical, renormalized 
 conductivity ${\bf \sigma_{xx}}$. 
Nevertheless one can use the phase diagram of this model to 
draw qualitative conclusions about physical conductivities. First of all, we
 have the striking fact that  the beta function of the WZNW1-
 model is parabolic at the critical point:
\[
\beta(g) \sim g^2
\]
This means that there is some critical amount of disorder,
 beyond which the
function $C(\omega)$ becomes insensitive to the boundaries in large samples 
even at $\sigma_{xy} = 1/2$. It is the most likely that this corresponds 
to localization. 

 Then,  the beta-function of the {\it supersymmetric} sigma model,
 which generates the physical observables\cite{weiden}, 
 should have  the same quadratic behaviour at the critical point.
Such conjecture was first made by Polyakov
\cite{pol}. 
 Although we can not deduce 
 the exact value of the critical conductivity $ \sigma^* ( E) $
 as a function of the scattering rate $ 1/\tau$,
 we do conclude  
 the existence of a Kosterlitz- Thouless transition as a function 
 of the disorder strength.

 It is believed that deviations of $\sigma_{xy}$ from 1/2 are  relevant,  
leading to a finite correlation length 
$\xi \sim (\sigma_{xy} - 1/2)^{-\nu}$ with $\nu \approx 7/3$
( see \cite{hucke} for a review).
 In addition there is numerical \cite{weng}
 and experimental evidence\cite{shahar},
 and it was conjectured on phenomenological grounds\cite{hajdu}  
 that there is a semi-elliptic separatrix in 
 the flow diagram which separates the  
 quantum Hall state from the insulating state.
   According to \cite{weng}, \cite{shahar}
 the Hall resistivity is quantized to $\rho_{xy} =
 1$ along this separatrix, yielding a semi circle
 defined by $ \sigma_{xx}^2 + ( \sigma_{xy} - 1/2 )^2 = 1/4$.
 Additional evidence for such a flow diagram comes from bifurcation theory:
  If the beta function would cross zero, there would be a repulsive node 
 and a saddle point on the $\sigma_{xy} = 1/2 $- line.
  Demanding that the beta function only touches zero, which is the
 rigourous result obtained from the O(3) NL sigma model,
 these two points merge and one has a fold- Hopf bifurcation with one 
 separatrix.   
 Therefore we  conjecture the following 
 set of RG equations for the real conductances 
in the vicinity of the critical point:
\begin{equation}
\frac{\partial \epsilon_{n} }{\partial\ln L} = - \epsilon_{n}^2.
\end{equation}
 where $\epsilon_{n}$ is the 
amplitude of deviation of the vector
 $ ( \sigma_{xx}, \sigma_{xy} ) $ normal 
 to the separatrix defined by $ \rho_{xy} = 1$, 
 or $ \sigma_{xx}^2 + ( \sigma_{xy} - 1/2 )^2 = 1/4$. 
 And
\begin{equation}
\frac{\partial \epsilon_t}{\partial \ln L} = \frac{1}{\nu} 
 \epsilon_t
\end{equation}
where  $\epsilon_t $ denotes  the  change of that vector
  along the separatrix.
This system of equations generates the scaling portrait shown in Fig. 1.  

 \begin{figure}[bhp]\label{fig2}
\epsfxsize=9cm
\centerline{\epsffile{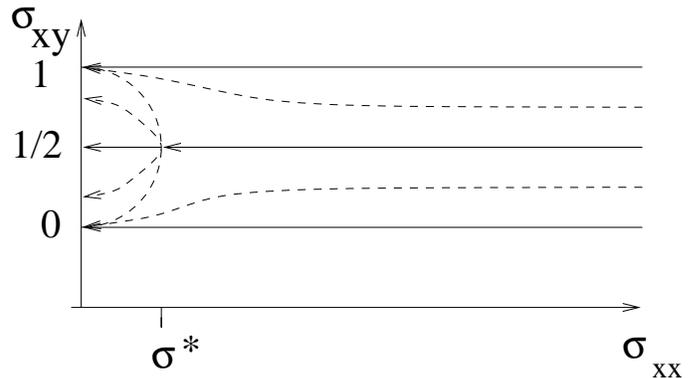}}
\caption{  The two- parameter flow diagram in the integer quantum hall regime.
 The full lines are based on the exact result. The broken lines
 are conjectured}  
\end{figure}    

  Recently, it has been shown that the SUSY (supersymmetric) 
NL sigma model  can be mapped 
 at $\sigma_{xy} =1/2~ mod~ (1) $ on
 the Hamiltonian of superspin chains\cite{annalen}. 

  It was argued that the SUSY model with the topological term 
does not have a classical critical point  in 
 two dimensions and therefore the critical point may appear only when
 it is  driven under renormalization 
 to  another theory. The results given above do
 suggest strongly   that the WZNW1- model is
  an important part of this theory. 

 We note that   the flow diagram, 
 as conjectured by
Khmelnitskii\cite{khm} seems inconsistent 
 at small bare longitudinal conductivities
with the one obtained above. 
 A derivation of the former  
 was given  by Pruisken based on a dilute instanton 
 approximation in the calculation of the beta- function of the 
 N- replica NL sigma model\cite{pruisken}.
 The validity of that derivation 
 is limited to large  conductivities, however,
 as noticed before  \cite{weiden,annalen,ef}. 
 In fact, for large bare conductivities both flow diagrams
 do agree with each other.
 
 The approach presented here can also be used  for  
 understanding  the influence of electron-electron  interactions 
 on the localization.
   In that case one studies the $n=1$- replica function,
 which is the autocorrelation function of the partition 
 function of the interacting electrons. 
 Considering only spinless fermions, 
  the electron - electron interaction breaks the  
 $U(2) $- invariance of the Grassmann fields.  
 Accordingly, as first derived  by A. Finkel'stein\cite{finkel}, decoupling the
 three interaction terms (one produced by averaging over impurities 
 which couples both partition functions and two real interaction 
 terms between the fermions in each partition function),
     and integrating out perturbatively the Hubbard- Stratonovich fields 
 originating from the real fermion interactions, leads 
 to the  O(3) NL sigma model action 
 where the  symmetry is broken by the 
 presence of the interactions.
 This results in an additional
 term of the form,
\begin{equation}
 \frac{U}{\Delta} \int \frac{d x^2}{L^2}  Tr [ Q({\bf x}) , \sigma_3 ]^2. 
\end{equation}      
 where $U$ is a spatially averaged measure of the 
  electron- electron interaction. 
 We note that the addition of this asymmetry can lead to the existence of 
  transitions in two dimensions:
 The O(2)  NL sigma model has  
 a Kosterlitz- Thouless transition, and 
   the Ising- model  has a phase transition in 2D.
   Thus, we can understand that the presence of interaction may indeed 
 result in the existence of a metal-insulator transition
 in two dimensions. 
  
 In summary  we have reconsidered the Anderson localization in two dimensions
 and the integer quantum Hall transition from the perspective of
 the level statistics. 

    For weak magnetic fields our results indicate that   
 all states are localized in the infinite volume limit.
   
   In  strong magnetic fields we have found that when  the
 bare  Hall conductance  is half integer $ \sigma_{xy}= 1/2,3/2,...$, 
 there do exist critical states at the corresponding energy. 
  The resulting flow diagram 
 shows a Kosterlitz- Thouless transition at a critical disorder strength  
 beyond which  there is no  extended state. 
  
 We note that  Quantum Hall- Insulator- transitions  
 have been recently observed experimentally 
 \cite{jiang}.   
 
 We acknowledge the kind hospitality of ICTP in Trieste where this work was completed. 
We are also grateful to B. Altshuler, A. Altland, K. Efetov, V. Fal'ko, 
I. Kravtsov, I. Lerner and  B. Simons for many valuable discussions 
and interest to the work. 
 S. K.  is particularly grateful to 
 P.  Fulde for his interest and support of this work and 
 acknowledges useful discussions with 
 M. R.Zirnbauer,  S. Villain-Guillot and G. Jug.
 
\end{document}